# Analysis of Possible Attack on AODV Protocol in MANET

Nitesh Funde[#1], P. R. Pardhi[*2]

[#] *M Tech Scholar,* [*] *Assistant Professor,*
*Department Of Computer Science & Enggineeing*
*Shri Ramdeobaba College Of Engg. & Management  Nagpur, India*

*Abstract—* **Mobile Ad Hoc Networks (MANETs) consist of wireless mobile nodes which coordinate with each other to form temporary network without its pre-existing infrastructure. AODV is popular Ad-hoc distance vector routing reactive protocol which is used to find correct & shortest route to destination. Due to openness, dynamic, infrastructure-less nature, MANET are vulnerable to various attacks. One of these possible attacks is a Black Hole Attack in which a mobile node falsely replies to the source node that it is having a shortest path to the destination without checking its routing table. Therefore source node send all of its data to the black hole node and it deprives all the traffic of the source node. In this paper, We are proposing a technique to detect and prevent the multiple black hole nodes from MANET so that source to destination communication can be made easily. We also analysed the performance of the network in terms of number of packets sent, received, throughput, energy of network before attack and after detection & prevention of Attack. From these analysis, we can conclude that performance decreased due to attack can be improved after detection & prevention black hole attack in MANET.**

*Keywords—* **AODV, black hole, RREQ, RREP, RERR**

## I. INTRODUCTION

Mobile Ad Hoc Networks (MANETs) is a self-configuring network of wireless mobile nodes that formed network capable of dynamic changing topology. Each node in the network acts as a router, forwarding data packets to other nodes [1]. MANET have many potential applications such as military services in battlefield, disaster relief operations and in commercial environments.

Routing in MANET is complex due to its mobility of nodes and dynamic changing topology as compared to traditional wired networks. Limited bandwidth and battery makes routing in MANET more challenging. Due to these fundamental characteristics of MANET, it is susceptible to various kinds of attacks like eaves dropping with malicious intent, spoofing of control or data packets, malicious modification of the packet contents and Denial of service attack like worm hole, sink hole, black and gray hole attacks [7]. These are all network layer attacks. So routing security is one of the important issue for which researchers want to contribute. Routing protocol plays vital role in security of the network. AODV is routing protocol designed and used for MANET to establish route on demand. It does not need to maintain routes which are not active.

In this paper, We attempt to provides a solution to detect the multiple black hole nodes present and prevent them from the network. In particular , we are focusing on AODV protocol in MANET .This solution are not only provide protection mechanism against black hole attack but also consequently improve the performance of the network comparing with the existing approaches after detection and prevention of attack. The analysis shows that how severe the attack is and its effects on MANET.

## II. BLACK HOLE ATTACK IN AODV PROTOCOL

Ad-hoc On-Demand Distance Vector (AODV) Routing Protocol is used for finding a path to the destination in an ad hoc network. To find the path to the destination all mobile nodes have to work in cooperation using the routing control messages. There are three types of routing control messages in AODV protocol Route Requests (RREQs), route Reply (RREP), Route Error (RERR) used to find a path to the destination [1]. The AODV routing protocol uses a destination sequence number for each route entry. The destination sequence number is generated by the destination when a connection is requested to it. The principle of this protocol is greater the destination sequence number, fresher is the route [1]. Small hop count is selected at the stage when most of the nodes have same retransmission time. When the source node S want to communicate with destination D as shown in the figure1,it broadcasts RREQs messages to the neighbour nodes. These neighbours check their routing table whether there is a path to the destination or not. If it is not then they also forward RREQs of the source node until message is received by the intermediate node having path to destination or destination node itself. If the node having a path to destination receives RREQ, it send route reply message to the source node. In this way, source node select the shortest path to the destination node with the greater sequence number of route reply message. If any link break occurs then RERR message send to source node.

In Black Hole Attack in MANET, here we assume malicious node M As Source S broadcasts its RREQs to connect with destination D, all intermediate node check their routing table whether there is a path to the destination D. Here Node M as malicious does not check its routing table and send false route reply packet to the source S with greater forged sequence number [3] than expected that it is having a path to destination D. As malicious node M does not check its route table, this reply reach the source node faster than the normal nodes [6] source node select this path and send all of its data to the node M. The node M receive this data packets and deprive from the destination node.As this packets never reach to the destination node D the attack is called as a black hole attack [2].

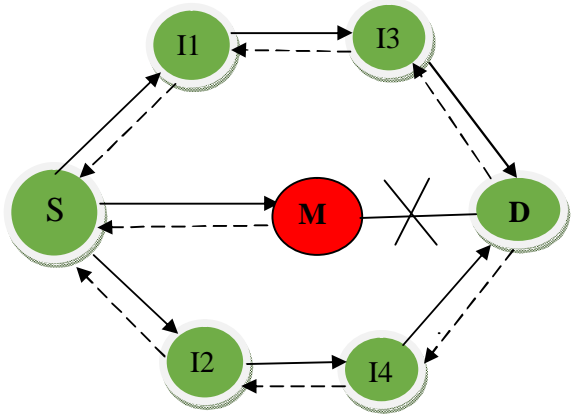

Fig1. Black Hole Attack

RREQ ⟶ RREP --→

### III. RELATED WORKS

Author [4] have discussed a approach where intermediate nodes have to send RREP message along with the next hop information. When the source node receive this information, it sends a route request to the next hop to verify that node which send back RREP packet has a route to the intermediate node and to the destination node. When the next hop receives a further request from the source node, it sends a further reply message which includes the check result to the source node that it is having a path or not. Based on this information, the source node judges the validity of the route. In this protocol, the RREP control packet is modified to contain the information about next hop. After receiving RREP, the source node will again send RREQ to the node specified as next hop in the received RREP. However, this increases the routing overhead and end-to-end delay.Here intermediate node have to send route reply packet twice for single RREQ.

In this, [5] author proposed a solution based on Intrusion Detection using Anomaly Detection (IDAD). It prevents both attacks by single and multiple black hole nodes. This mechanism assume that every activity of a user is monitored and abnormal activities of malicious node can be identified from normal activities of normal nodes.For detection of the black hole attack, this method having a pre collected set of anomalies activities ,abnormal activities called audit data. As soon as the audit data collected and given to the system it compares every activity with the audit data.If any activity is not listed in audit data , then particular malicious node would have to be isolated and remove from the network. But the drawback is that if neighbour node give false information then this solution lead to more delay in MANET.

### IV. PROPOSED APPROACH

Aim of the proposed algorithm is to detect multiple blackholes in MANET and improves performance of the network in terms of throughput, energy which degrades due to the attack. The proposed algorithm is consists of three main steps.

#### A. Initializing Source and Destination Node

Here first source node want to communicate with the destination. In AODV protocol, source node have a predetermined set of time for receiving RREP route reply packets from intermediate nodes or destination node itself. Source node analyse one by one.

#### B. Analysing RREP Route Reply Packets

The source node broadcast its RREQs to neighbours, when they have the path to destination, they reply to the source node about the path information which contains destination sequence number, next hop, destination node. According to principle of AODV protocol, greater the sequence number fresher is the route. Source node select path through the intermediate node having greater sequence number.

Normal node reply with the sequence number to the source node. We are expecting a threshold value of sequence number to be 1000 which is any how maximum for the network having less than 50 number of nodes. Here, We are creating scenario where 25 nodes present in the network. But as soon as malicious node receive route request from the source node, it just send false reply with greater sequence number as compared to normal nodes to get entry in the routing table of the source node with minimum time.

As source node get the reply from black hole node faster and greater sequence number, source node chooses this path and it leads to black hole attack. So we are considering the route reply adaptive sequence number of each intermediate node to the source node about the path information of destination and compare it with the threshold value which is considered to be maximum of the number of nodes present in the network. The adaptive sequence number is calculated by using reply message with the help of time keeper register.

#### C. Comparing the adaptive RREP sequence number with threshold value

When reply is having sequence number greater than the threshold value in a particular time duration, definitely it is black hole node. In this way, We are detecting multiple black hole nodes one by one and black listed them so that in the future, they may not be participate in node discovery process in the network.

### V. PSEUDO CODES

Step 1: Initialize the source and destination node in the network.
Step 2: Source Node broadcast its RREQs to communicate with the destination.
Step 3: Adding time register for calculating the adaptive sequence numbers of incoming RREPs to the source node.

Step 4: Comparing the adaptive destination sequence number with threshold value in a particular time duration.
if(RREP_SEQ_NO > THRESHOLD VALUE)
    Malicious Node = Node;
    BH Node Detected, So re-routing the packet;
else
    Select the normal path through which source can communicate with the destination node.
 End
Step 5: Select the secure route by choosing a intermediate node for transmission
Step 6: go to step 3.
Step 7: End.

VI. SIMULATION RESULTS & ANALYSIS

A. *Simulation Environment*

For simulation we have used NS-2 (ver-2.35) simulator on Ubuntu 13.04 operating system and analyze with the help of Tracegraph application and Xgraph AWK scripts.The simulation results involves network topology with 25 nodes where there are different source nodes communicates with the destination node. Mobility scenarios can be generated by random waypoint model in terrain area 1286m * 850m.The simulation parameters are as shown in table 1.

TABLE 1
SIMULATION PARAMETERS

| Parameters | Values |
|---|---|
| Simulator | NS-2 (ver-2.35) |
| Simulation Time | 20 sec |
| Number Of Nodes | 25 |
| Routing Protocol | AODV |
| Traffic Model | CBR |
| No of sources | 4 |
| Terrain Area | 1286m * 850m |
| Transmission Range | 250 m |
| No of malicious nodes | 3 |

B. *Simulation Analysis & Results*

We created scenario of 25 nodes MANET and divides it into 3 phases as MANET without Black hole attack, Blackhole attack on MANET and Detection & prevention of attack and simulated all this scenario with the help oNS-2 simulator. We configure different source nodes 21, 20, 11 17 to communicate with the destination node 18 at different time. Node 21 communication with the destination node 18 scenario black hole attack phase and after detection & prevention phase is as shown on simulation screenshots in figure 2 & 3.In figure 3, source node send the data packets to the black hole node 1 instead of destination node 18 through intermediate node 15.

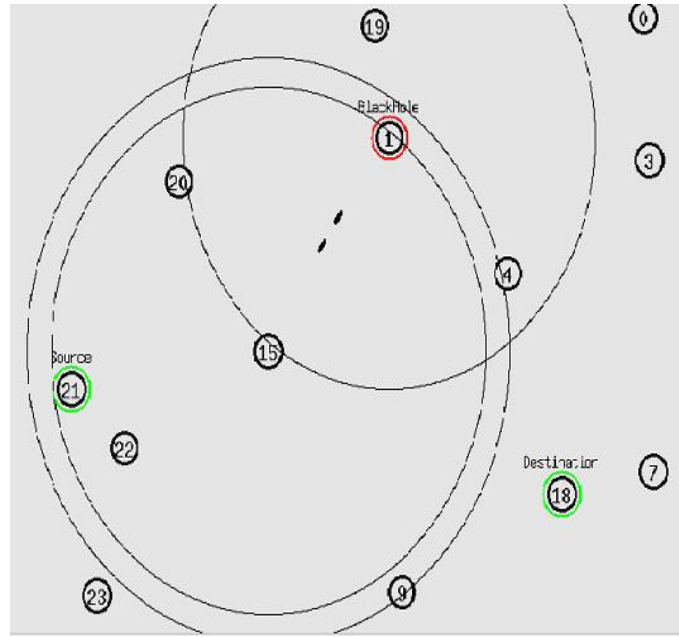

Fig.2 Black hole Attack (Source Node 21- Destination Node 18)

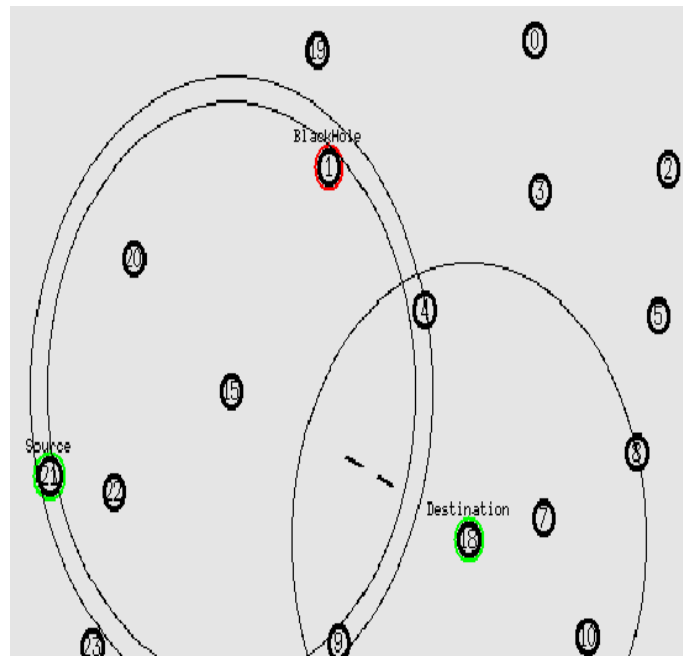

Fig.3 After Detection & Removal Of Attack (SN 21- DN 18)

The MANET without Black hole attack always performs better in terms of throughput and energy consumption. After Black hole attack, we also measure the number of packets sent by the source node, received by destination node by using is analysed with the help of Tracegraph application. In this , We compare and analyse the performance of MANET in all three phases. Our results shows that the performance degradation of MANET after black hole attack is improved in terms of throughput after detection and removal of multiple black hole nodes from MANET. The analysis in terms of performance comparison of all the three phase is as shown as below. In fig.4 The Number of receiving packets at destination node vs simulation time of the network graph is shown.

In this graph, we can analyse that no of received packets at destination from different source decreased due to black hole attack but after removal of attack it increases. Green line indicates MANET without black hole attack, red line indicates black hole attack and blue line indicates detection & removal of attack. The result shown in fig.5 shows that energy consumption is more when we apply our detection technique for black hole attack.

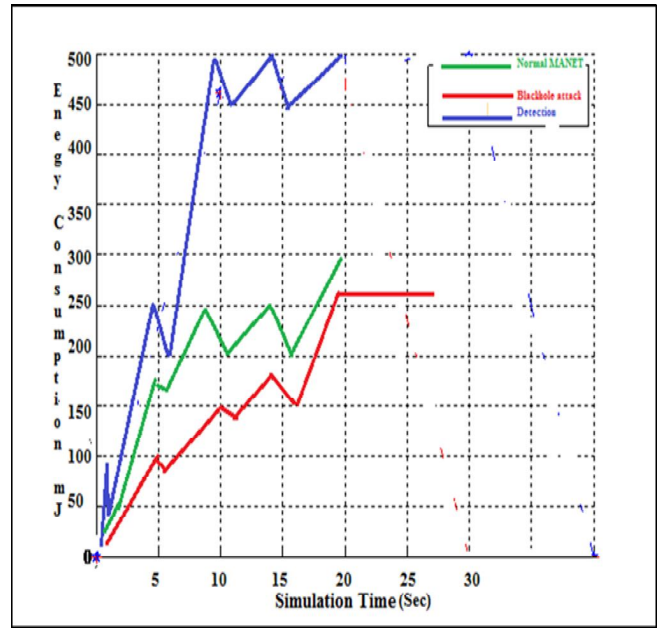

Fig.5 Energy Consumption of the Network

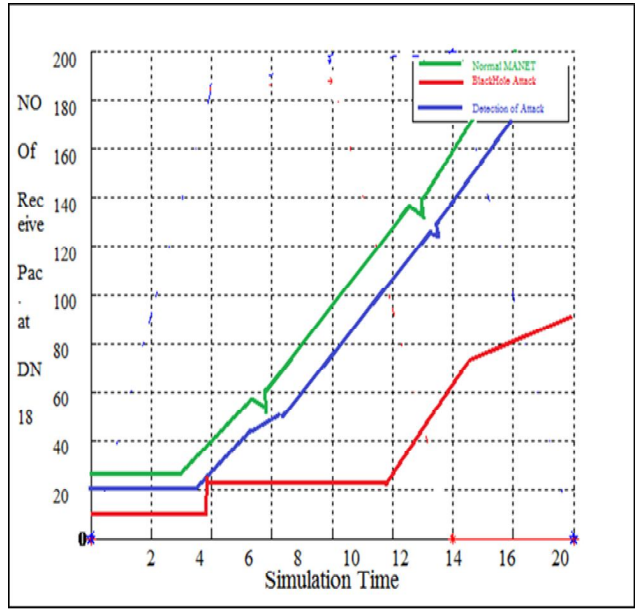

Fig.4. No Of Received Receiving Packets at Destination node

### VII. CONCLUSIONS & FUTURE WORK

The simulation analysis and results showed that black hole attack is one of the possible severe attack that can be easily launched on MANET. Hence, the proposed algorithm to detect black hole attack not only correctly detect the multiple black hole nodes in MANET but also improve the performance of the network after removal of attack. Our proposed algorithm performs better in terms of throughput of the network with some increased in energy consumption of the network. So in future, we will try to improve algorithm in terms of minimum energy consumption. We have carried comparative analysis of different phases which shows that black hole attack effect on the performance of the network.
Nodes mobility can effect on the performance of MANET in terms of misdetection probability. We can considered this as a false positives and in future we will calculate misdetection probability based on mobility of nodes in MANET. We also emphasize that though the proposed algorithm is implemented and simulated for the AODV routing algorithm, it can also be further extended for use by any other routing algorithms, as well.

### ACKNOWLEDGMENT

I express my sincere gratitude to Prof. P. R. Pardhi Department of CSE, for his valuable guidance and advice. Also I would like to thanks Dr. M. B. Chandak Head Department of CSE for continuous guidance, support and encouragement.